\documentclass[numreferences]{kluwer}
\let\iint=\undefined

\usepackage{amsmath,graphicx}

\newcommand{\dee}{\partial}
\def\xvec{{\bf x}}
\def\bvec{{\bf v}}
\def\uvec{{\bf u}}

\begin{document}

\begin{opening}

\title{Stable simulation of fluid flow with high-Reynolds \\
number using Ehrenfests' steps}
\author{R. \surname{Brownlee}, A. N.  \surname{Gorban}, and J.  \surname{Levesley}}
\institute{Department of Mathematics, University of Leicester,
Leicester LE1 7RH, UK}
\runningauthor{R. Brownlee, A. N. Gorban, and
J. Levesley}
\runningtitle{Stable simulation of fluid flow using
Ehrenfests' steps}
\date{2nd March 2007}

\begin{abstract}
The Navier--Stokes equations arise naturally as a result of
Ehrenfests' coarse-graining in phase space after a period of
free-flight dynamics. This point of view allows for a very flexible
approach to the simulation of fluid flow for high-Reynolds number.
We construct regularisers for lattice Boltzmann computational
models. These regularisers are based on Ehrenfests' coarse-graining
idea and could be applied to schemes with either entropic or
non-entropic quasiequilibria. We give a numerical scheme which gives
good results for the standard test cases of the shock tube and the
flow past a square cylinder.
\end{abstract}


\end{opening}

\section{Introduction}

The simulation of high-Reynolds number flow is notoriously
difficult. In two space dimensions, a partial differential equation
model for such flows is the Navier--Stokes equations:
\begin{equation}
\begin{split}
{\partial \rho \over \partial t} & =  -\nabla \cdot (\rho \uvec),  \\
{\partial \over \partial t} ( \rho u_1) & =  -\sum_{j=1}^2 {\partial
 \over \partial x_j} (\rho u_1 u_j)  - {\partial P \over
\partial x_1}   \\
& \qquad + \mu \biggl( {\partial \over \partial x_1} P \biggl(
{\partial u_1 \over \partial x_1} -  {\partial u_2 \over \partial
x_2} \biggr)
+ {\partial \over \partial x_2} P \biggl(  {\partial u_2 \over \partial x_1} +  {\partial u_1 \over \partial x_2} \biggr) \biggr),  \\
{\partial \over \partial t} ( \rho u_2) & =  -\sum_{j=1}^2 {\partial
 \over \partial x_j} (\rho u_2 u_j)  - {\partial P \over
\partial x_2} \\
& \qquad + \mu \biggl( {\partial \over \partial x_2} P \biggl(
{\partial u_2 \over \partial x_2} -  {\partial u_1 \over \partial
x_1} \biggr) + {\partial \over \partial x_1} P \biggl(  {\partial
u_1 \over
\partial x_2}
+{\partial u_2 \over \partial x_1} \biggr) \biggr),  \\
 \frac{\partial E}{\partial t} &= - \sum_{i=1}^2 \frac{\partial}{\partial x_i}  \Bigl\{  u_i ( E+P ) \Bigr\} +
 {\mu} \sum_{i=1}^2  \frac{\partial}{\partial x_i} \biggl( P  \frac{\partial P}{\partial x_i}  \biggr),
\end{split}\label{nseq}
\end{equation}
where $\rho$, $\uvec=(u_1,u_2)$, $P$ and $E$ are density, velocity,
pressure and energy density respectively. These equations model the
conservation of mass, momentum and energy. The number $\mu$ is the
coefficient of viscosity, and as this number tends to zero we
recover the Euler equations for inviscid flow. The Reynolds number
of the flow is
\begin{equation*}
  \mathrm{Re}  = \frac{L u_\infty}{\nu},
\end{equation*}
where $L$ is the characteristic length scale in the problem,
$u_\infty$ is the free-stream fluid velocity, and $\nu=\mu/\rho$ is
the kinematic viscosity. As $\mu \rightarrow 0$, $\mathrm{Re}
\rightarrow \infty$.

Our aim is to model the flow in a physical way, so that the limit as
the viscosity gets small is the Euler equations, but also that
diffusion is added in a targeted, physical and controlled way. We
will present a variant of the lattice Boltzmann method which was
introduced in~\cite{Rob}.  We will also set this method in the
context of a more general coarse-graining paradigm.

\section{The lattice Boltzmann method}\label{sec2}

Let $f=f(\xvec,\bvec,t)$ be a one-particle distribution function,
i.e., the probability of finding a particle in a volume
$\mathrm{d}V$ around a point $(\xvec,\bvec)$, at a time $t$, in
phase space is $f(\xvec,\bvec,t)\mathrm{d}V$. Then, Boltzmann's
kinetic transport equation is the following time evolution equation
for $f$,
\begin{equation}
    \frac{\partial f}{\partial t}+\bvec \cdot \nabla f  =  Q(f).\label{be}
\end{equation}
The \emph{collision integral}, $Q$, describes the interactions of
the \emph{populations} $f$.

Equation~\eqref{be} describes the microscopic dynamics of our model.
We will wish to recover the macroscopic dynamics, the fluid density,
momentum density and energy density.

We do this by integrating the distribution function:
\begin{align*}
\rho(\xvec,t) & :=  \int f(\xvec,\bvec,t)\,\mathrm{d}\bvec, \\
\rho u_i (\xvec,t) & :=  \int v_i f(\xvec,\bvec,t) \,\mathrm{d}\bvec, \qquad \text{$i=1,2$,} \\
E(\xvec,t) & :=  {1 \over 2} \int \bvec^2 f(\xvec,\bvec,t)
\,\mathrm{d}\bvec.
\end{align*}
Such functionals of the distribution are called \emph{moments}. The
pressure $P$ is given by
\begin{equation*}
    P =  E - \frac{1}{2} \rho \uvec^2,
\end{equation*}
where we have set Boltzmann's constant to 1.

The lattice Boltzmann approach drastically simplifies this model by
stipulating that populations can only move with a finite number of
velocities $\{\bvec_1,\dotsc,\bvec_n\}$:
\begin{equation}\label{lbe}
    \frac{\partial f_i}{\partial t}+\bvec_i \cdot \nabla f_i  =
    Q_i,\qquad \text{$i=1,\dotsc,n$,}
\end{equation}
where $f_i$ is the one-particle distribution function associated
with motion in the $i$th direction.

Let $m$ be the linear mapping which takes us from the microscopic
variables $f$ to the vector of macroscopic variables $M$:
\begin{equation*}
M := (\rho,\rho u_1,\rho u_2,2E).
\end{equation*}
There are an infinite number of distribution functions which give
rise to any particular macroscopic configuration $M$. Given a
strictly concave entropy functional $S(f)$, for any fixed $M$ there
will be a unique $f$ which is the solution of the optimisation
\begin{equation}\label{argmax}
f^*_M=\arg\max\Bigl\{ S(f) : m(f)=M \Bigr\}.
\end{equation}
We call $f^*_M$ the \emph{quasiequilibrium} as it is not a global
equilibrium. The manifold of quasiequilibria, parameterised by the
macroscopic moments $M$, is called the \emph{quasiequilibrium
manifold}.

If the entropy is the Boltzmann entropy
\begin{equation*}
S(f) = -\iint f \log f\,\mathrm{d}\bvec\mathrm{d}\xvec
\end{equation*}
the quasiequilibrium is the Maxwellian distribution
\begin{equation}\label{max}
  f^*_M(\bvec) =  \frac{\rho^2}{2 \pi P} \exp \biggl( -{\rho
\over P} (\bvec-\uvec)^2 \biggr).
\end{equation}

Since $m(f)=m(f^*_M)$ an integration rule which evaluates
\begin{equation*}
  \int g(\bvec) f^*_M(\bvec)\,\mathrm{d}\bvec\
\end{equation*}
exactly for low-degree polynomials will preserve the conservation of
the macroscopic variables $M$. Since the Maxwellian~\eqref{max} is
essentially Gaussian, the first candidate for this is a
Gauss--Hermite-type integration formula. If we do this we get an
approximation
\begin{equation*}
\int g(\bvec) f(\bvec)\,\mathrm{d}\bvec \approx \sum_{i} W_i
g(\bvec_i) f(\bvec_i).
\end{equation*}
If we write $f_i(\xvec)= f(\bvec_i)$ then we can view the lattice
Boltzmann equation~\eqref{lbe} as a quadrature approximation in the
velocity variable to Boltzmann's equation~\eqref{be}. For a complete
treatment of this point of view see~\cite{shanhe}.

In the lattice Boltzmann community the collision of choice has
become the Bhatnager--Gross--Krook~\cite{bgk} collision
\begin{equation}\label{bgkcoll}
Q(f,f)=-\omega(f-f^*).
\end{equation}
This is due to its simplicity and the nice intuitive interpretation
that the dynamics relaxes towards the quasiequilibrium in a time
that is proportional to the relaxation time $\tau=1/\omega$, which
models viscous processes. Via the Chapman--Enskog procedure it can
be shown (see, e.g.,~\cite{succi01}) that the associated macroscopic
dynamics is the Navier--Stokes equations to second-order in $\tau$.

The kinetic equation appears as an intermediate object between the
macroscopic transport equations and the numerical LBM simulation.
But, unfortunately, in the very intriguing limit of small viscosity
and time step $\Delta t > \tau$, the discrete LBM model can not be a
good approximation for the continuous-in-time kinetics.
Nevertheless, the discrete model can still provide $\Delta t^2$
accurate approximation of the macroscopic transport
equations~\cite{BGJPreprint,BGJ}.

We will see that it is possible to avoid the use of the kinetic
equation as an intermediary between LBM and the hydrodynamic
context. What we will demonstrate here (following the method
presented in~\cite{GorKar,GKOeTPRE2001}) is that the Navier--Stokes
equations arise in a natural way via free-flight dynamics for a time
$\tau$, followed by equilibration. The coefficient of viscosity will
be $\tau/2$. For the remainder of the paper, the
\emph{coarse-graining time}, $\tau$, should not be confused with the
relaxation time in~\eqref{bgkcoll}. We will also show that we can
approximate the Euler equations to order $\tau^2$ by a judicious
choice of numerical scheme. A more detailed treatment of the
construction of such numerical schemes for the approximation of the
Navier--Stokes and Euler equations may be found
in~\cite{Rob2,BGJPreprint,BGJ}.

\section{Coarse-graining}

The original Ehrenfests' method~\cite{ehrenfest11} for introducing
diffusion into a system was to divide phase space into cells. Then,
after the dynamic motion of the microscopic ensemble under
\eqref{be}, which is conservative, an averaging occurs in the cells,
giving rise to an entropy increase. Many other methods in
statistical mechanics can be understood as a generalisation of this
coarse-graining paradigm~\cite{gorban06}. We will be using a
modified lattice Boltzmann method to simulate the flow and we will
describe this in the more general context of coarse-graining.

We start from the phase flow transformation of the conservative
dynamics: $\Theta_{\tau}:f(\xvec,t) \mapsto f(\xvec,t+\tau)$. For
the Ehrenfests' this was the flow of the Liouville equation. We
mostly use the free-flight conservative dynamics:
$\Theta_{\tau}:f(\xvec,\bvec,t) \mapsto f(\xvec-\bvec\tau,\bvec,t)$
(this means: $f(\xvec,\bvec,t+\tau)=f(\xvec-\bvec\tau,\bvec,t)$).

Let $\tau$ be a fixed coarse-graining time and suppose we have an
initial quasiequilibrium distribution $f_0$. The Ehrenfests' chain
$f_0, f_1,\dotsc$ is the following sequence of quasiequilibrium
distributions:
\begin{equation*}
  f_{i}  = f^*_{m(\Theta_\tau(f_{i-1}))},\qquad
  \text{$i=1,2,\dotsc$.}
\end{equation*}

Entropy increases in the Ehrenfests' chain. By virtue of the
conservative dynamics there is no entropy gain from the mechanical
motion (from $f_i$ to $\Theta_\tau(f_i)$), the gain follows from the
equilibration (from $\Theta_\tau(f_i)$ to $f_{i+1}$). Consequently,
conservative systems become dissipative.

\subsection{Determining macroscopic dynamics}

We wish to determine the macroscopic dynamics which passes through
the points $m(f_i)$, $i=0,1,\dotsc$. In general, this will depend on
the parameter $\tau$, so we seek an equation of the form
\begin{equation*}
\frac{\dee M}{\dee t} = F(M,\tau).
\end{equation*}
Following~\cite{GorKar,GKOeTPRE2001} we will expand this for small
$\tau$ in a series $F(M,\tau)=F_0(M)+\tau
F_1(M)+\mathcal{O}(\tau^2)$ and match terms in powers of $\tau$ to
determine $F_0$ and $F_1$. In other words, for any quasiequilibrium
state $f_0$, we wish to have
\begin{equation*}
 m(\Theta_\tau(f_0)) = M(\tau)
\end{equation*}
to second-order in $\tau$.

In phase space we have chosen free-flight dynamics
\begin{equation*}
{\dee f \over \dee t} +\bvec \cdot \nabla f =0,
\end{equation*}
with exact solution
\begin{equation*}
\Theta_t(f_0)(\xvec,\bvec)=f_0(\xvec-\bvec t,\bvec).
\end{equation*}
Since $f_0$ is on the quasiequilibrium manifold we will replace
$f_0$ with $f^*$ from now on.

The second-order expansion in time for the dynamics of the
distribution $f$ is
\begin{align*}
    \Theta_\tau(f^*) &=  \Theta_0(f^*)+ \tau {\dee \Theta_t \over \dee t} \biggr|_{t=0} + {\tau^2 \over 2}  {\dee^2 \Theta_t \over \dee t^2} \biggr|_{t=0} \\
    & =  f^* - \tau \bvec \cdot \nabla f^* +  {\tau^2 \over 2}  \bvec \cdot \nabla (  \bvec \cdot \nabla f^* ).
\end{align*}
Thus, to second-order,
\begin{equation*}
m(\Theta_\tau(f^*)) = m(\Theta_0(f^*)) - \tau  {\dee  \over \dee t}
m (\bvec \cdot \nabla f^*) +  {\tau^2 \over 2} m ( \bvec \cdot
\nabla (  \bvec \cdot \nabla f^* )).
\end{equation*}

Similarly, to second-order,
\begin{align*}
M(\tau) & =  M(0)+\tau  {\dee M  \over \dee t} \biggr|_{t=0} + {\tau^2 \over 2}  {\dee^2 M \over \dee t^2} \biggr|_{t=0} \\
& =  M(0)+\tau ( F_0(M)+\tau F_1(M) ) + {\tau^2 \over 2} {\dee
F_0(M) \over \dee t}.
\end{align*}
Since $M(0)= m(\Theta_0(f^*))$, we have
\begin{multline*}
- \tau  m ( \bvec \cdot \nabla f^*) + {\tau^2 \over 2} m ( \bvec
\cdot \nabla (  \bvec \cdot \nabla f^* ))  \\ =\tau ( F_0(M)+\tau
F_1(M) ) + {\tau^2 \over 2} {\dee F_0(M) \over \dee t}.
\end{multline*}

Comparing the first-order terms we have
\begin{equation*}
 F_0(M)  = -m (  \bvec \cdot \nabla f^* ).
\end{equation*}
Comparing the second-order terms gives
\begin{equation*}
F_1(M)+{1 \over 2} {\dee F_0(M) \over \dee t}  =  {1 \over 2} m (
\bvec \cdot \nabla ( \bvec \cdot \nabla f^* )),
\end{equation*}
and, upon rearrangement, we get
\begin{equation*}
F_1(M) = {1 \over 2} \biggl( m ( \bvec \cdot \nabla (  \bvec \cdot
\nabla f^* )) - {\dee F_0(M) \over \dee t} \biggr).
\end{equation*}
Hence, to second-order, the macroscopic equations are
\begin{equation*}
{\dee \over \dee t} m(f^*) =  -m ( \bvec \cdot \nabla f^* ) + {\tau
\over 2} \biggl( m ( \bvec \cdot \nabla ( \bvec \cdot \nabla f^* ))
- {\dee F_0(M) \over \dee t} \biggr).
\end{equation*}

In what follows, to aid the flow of the presentation, we consign
some of the calculations to an appendix. Now, let us look at the
term $m(\bvec \cdot \nabla f^*)$ more carefully. The first component
is
\begin{equation}\label{mass}
\begin{split}
m_1( \bvec \cdot \nabla f^*) & =  \int  \bvec \cdot \nabla f^*\,\mathrm{d}\bvec  \\
& =  \int v_1 {\dee f^* \over \dee x_1} +  v_2 {\dee f^* \over \dee x_2}\,\mathrm{d}\bvec  \\
& =   {\dee \over \dee x_1} \int v_1  f^* \,\mathrm{d}\bvec +  {\dee \over \dee x_2} \int v_2  f^*\,\mathrm{d}\bvec  \\
& =   {\dee \over \dee x_1}( \rho u_1) +  {\dee \over \dee x_2} (\rho u_1)  \\
& =  \nabla \cdot (\rho u).
\end{split}
\end{equation}

Using~\eqref{ivivj}, the second component is
\begin{equation}
\begin{split}
m_2( v \cdot \nabla f^*) & =   \int  v_1 \bvec \cdot \nabla f^* \,\mathrm{d}\bvec \\
& =   {\dee \over \dee x_1} \int v_1^2  f^* \,\mathrm{d}\bvec +  {\dee \over \dee x_2} \int v_1 v_2  f^* \,\mathrm{d}\bvec  \\
& =   {\dee \over \dee x_1} \Bigl( {P} + \rho u_1^2 \Bigr) + {\dee
\over \dee x_2}  \rho u_1 u_2.
\end{split}
\end{equation}
Similarly,
\begin{equation}
\begin{split}
m_3( v \cdot \nabla f^*)  & =   \int  v_2 \bvec \cdot \nabla f^* \,\mathrm{d}\bvec  \\
& =   {\dee \over \dee x_1} \rho u_1 u_2 +  {\dee \over \dee x_2}
\Bigl( {P} + \rho u_2^2 \Bigr).
\end{split}
\end{equation}

Finally, using~\eqref{ivlvivj},
\begin{equation}\label{en}
\begin{split}
m_4(v \cdot \nabla f^*) & =   \int  \bvec^2 v \cdot \nabla f^* \,\mathrm{d}\bvec  \\
& =  {\dee \over \dee x_1} \int \bvec^2 v_1  f^* \,\mathrm{d}\bvec +  {\dee \over \dee x_2} \int \bvec^2 v_2  f^* \,\mathrm{d}\bvec  \\
& =   {\dee \over \dee x_1} \biggl( 4u_1{P} + \rho u_1 \uvec^2 \biggr) +  {\dee \over \dee x_2} \biggl( 4u_2{P} + \rho u_2 \uvec^2 \biggr )  \\
& =  2 \biggl(  {\dee \over \dee x_1} \Bigl\{ u_1 ( E+{P}) \Bigr\} +
{\dee \over \dee x_2} \Bigl\{ u_2( E+{P} ) \Bigr \} \biggr).
\end{split}
\end{equation}

Hence, from~\eqref{mass}--\eqref{en}, the first-order approximation
of the macroscopic dynamics is~\eqref{nseq} with $\mu = 0$, i.e.,
the Euler equations. More specifically we have (spelling out the
details for the first two components),
\begin{align}
F_{0,0}(M) & =   -{\partial \over \partial x_1} M_1  -{\partial \over \partial x_2} M_2,  \label{f00} \\
F_{0,1}(M) & =    -{1 \over 2} {\partial  \over \partial x_1} \biggl( M_3 - {M_1^2+M_2^2 \over M_0} \biggr) - {\partial  \over \partial x_i} {M_1 M_i \over M_0}, \label{f01} \\
F_{0,2}(M) & =    -{1 \over 2} {\partial  \over \partial x_2} \biggl( M_3 - {M_1^2+M_2^2 \over M_0} \biggr) - {\partial  \over \partial x_i} {M_2 M_i \over M_0}, \nonumber \\
F_{0,3}(M) & =   2 \biggl(  {\dee \over \dee x_1} \biggl\{  {M_1 M_3 \over M_0} +{M_1(M_1^2+M_2^2) \over M_0^2} \biggr\}  \nonumber \\
& \qquad\qquad\qquad +  {\dee \over \dee x_2} \biggl\{ {M_2 M_3
\over M_0} +{M_2(M_1^2+M_2^2) \over M_0^2} \biggr\} \biggr).
\nonumber
\end{align}

We now look at the second-order correction
\begin{equation*}
{1 \over 2} \biggl( m ( \bvec \cdot \nabla (  \bvec \cdot \nabla f^*
)) - {\dee F_0(M) \over \dee t} \biggr).
\end{equation*}
Due to the computational complication of what is to follow we will
only look at the first and second components of the above vector.
This will give the reader a flavour of the computation.

Let us look at the first component. Firstly,
\begin{equation*}
\begin{split}
 m_0 (  \bvec \cdot \nabla (  \bvec \cdot \nabla f^* )) & = \int\bvec \cdot \nabla (  \bvec \cdot \nabla f^* )  \,\mathrm{d}\bvec   \\
& = \sum_{i=1}^2 \sum_{j=1}^2 {\partial^2 \over \partial x_i \partial x_j} \int v_i v_j f^* \,\mathrm{d}\bvec  \\
& =  {\partial^2 \over \partial x_1^2} \Bigl\{ {P} + \rho u_1^2
\Bigr\} \\
&\qquad+ 2  {\partial^2 \over \partial x_1 \partial x_2} (\rho u_1
u_2) + {\partial^2 \over \partial x_2^2} \Bigl\{ {P} + \rho u_2^2
\Bigr\}.
\end{split}
\end{equation*}
Now, from~\eqref{f00},
\begin{equation*}
{\dee F_{0,0}(M) \over \dee t} =   \sum_{i=0}^3  { \dee F_{0,0}
\over \partial M_i} \biggl( {\partial M_i \over \partial t} \biggr)
 = \sum_{i=0}^3  { \dee F_{0,0} \over \partial M_i} F_{0,i},
\end{equation*}
where ${\dee F_0 \over \partial M_i}$ is viewed as an operator. Now,
\begin{equation*}
 { \dee F_{0,0} \over \partial M_0} =  { \dee F_{0,0} \over \partial M_3}  =
 0,\qquad  { \dee F_{0,0} \over \partial M_1} =   -{ \dee  \over \dee
 x_1}, \qquad  { \dee F_{0,0} \over \partial M_2}  =   -{ \dee  \over \dee x_2},
\end{equation*}
so that
\begin{align*}
{\dee F_{0,0}(M) \over \dee t} & =  -{ \dee  \over \dee x_1}
\biggl\{
 {\dee \over \dee x_1} \Bigl( {P} + \rho u_1^2 \Bigr) +  {\dee \over \dee x_2}  \rho u_1 u_2 \biggr\} \\
& \qquad\qquad - { \dee  \over \dee x_2} \biggl\{ {\dee \over \dee x_2} \Bigl( {P} + \rho u_2^2 \Bigr) +  {\dee \over \dee x_1}  \rho u_1 u_2 \biggr\} \\
& =   {\partial^2 \over \partial x_1^2} \Bigl\{ {P} + \rho u_1^2
\Bigr\} + 2  {\partial^2 \over \partial x_1 \partial x_2} (\rho u_1
u_2) +  {\partial^2 \over \partial x_2^2} \Bigl\{ {P} + \rho u_2^2
\Bigr\}.
\end{align*}
Hence, the first component of the second-order correction is zero,
which we would expect as this is the equation of mass conservation.

Now we look at the second term in the second-order correction, and
focus only on the pressure terms.

To compute the correction terms for the Navier-Stokes equations we
need to compute
\begin{equation*}
{\partial F_{0,1} \over \partial M_i}, \qquad \text{$i=0,1,2,3$.}
\end{equation*}
Now, we have from~\eqref{f01},
\begin{equation*}
F_{0,1} = -{1 \over 2} {\partial  \over \partial x_1} \biggl( M_3 -
{M_1^2+M_2^2 \over M_0} \biggr) - {\partial  \over
\partial x_i} {M_1 M_i \over M_0}.
\end{equation*}
Thus,
\begin{align*}
{\partial F_{0,1} \over \partial M_0} & =  -{1 \over 2}{\partial \over \partial x_1} u^2 + {\partial \over \partial x_i} u_1 u_i, \\
{\partial F_{0,1} \over \partial M_1} & =  - {\partial \over \partial x_1} u_1 -  {\partial \over \partial x_2} u_2, \\
{\partial F_{0,1} \over \partial M_2} & =  {\partial \over \partial x_1} u_2 - {\partial \over \partial x_2} u_1, \\
{\partial F_{0,1} \over \partial M_3} & =  -{1 \over 2}  {\partial
\over \partial x_1}.
\end{align*}
Hence,
\begin{align}
{\partial F_{0,1} \over \partial M_0}  F_{0,0} & = \biggl\{ -{\partial \over \partial x_1} u_2^2+{\partial \over \partial x_i} u_1 u_2 \biggr\} \biggl \{ -{\partial \over \partial x_i} \rho u_i \biggr \}, \label{t1}\\
{\partial F_{0,1} \over \partial M_1}  F_{0,1}  & =  - \biggl\{ {\partial \over \partial x_1} u_1 +  {\partial \over \partial x_2} u_2 \biggr \} {\partial \over \partial x_2} u_2 \biggl\{ -{\partial \over \partial x_1} {P} - {\partial \over \partial x_i} (\rho u_1 u_i) \biggr\}, \\
{\partial F_{0,1} \over \partial M_2}  F_{0,2} & =  \biggl\{  {\partial \over \partial x_1} u_2 -  {\partial \over \partial x_2} u_1 \biggr \} \biggl \{ -{\partial \over \partial x_2} {P} - {\partial \over \partial x_i} (\rho u_2 u_i) \biggr \}, \\
{\partial F_{0,1} \over \partial M_3}  F_{0,3} & =  -{1 \over
2}{\partial \over \partial x_1} \biggl\{ - 2 {\partial \over
\partial x_i} u_i \biggl(  {2P} + {1 \over 2} \rho \uvec^2
\biggr) \biggr \} \label{t2}. \end{align}

On the other hand, from~\eqref{ivlvivj}, we have
\begin{equation}\label{t3}
\begin{split}
{\partial^2 \over \partial x_i \partial x_j} \int v_1 v_i v_j f^* \,\mathrm{d}\bvec &=  {\partial^2 \over \partial x_i \partial x_j} \biggl\{ \Bigl( u_i+u_j+\delta_{i,j} u_1\Bigr){P} + \rho u_1 u_i u_j \biggr\} \\
&= 3 {\partial^2 \over \partial x_1^2} ( u_1 {P}) + {\partial^2 \over \partial x_1 \partial x_2} ( 2 u_2 {P} ) \\
& \qquad + {\partial^2 \over \partial x_2^2}  ( u_1 {P})+
{\partial^2 \over \partial x_i \partial x_j} (\rho u_1 u_i u_j) .
\end{split}
\end{equation}

If we denote by $D$ the difference of the pressure terms
in~\eqref{t3} and those in equations~\eqref{t1}--\eqref{t2} then we
obtain
\begin{align*}
D & =   3 {\partial^2 \over \partial x_1^2} ( u_1 {P} ) + {\partial^2 \over \partial x_1 \partial x_2} ( 2u_2 {P}  )+ {\partial^2 \over \partial x_2^2}  ( u_1 {P} ) \\
& \qquad - {\partial \over \partial x_1} u_1 \biggl\{ {\partial \over \partial x_1} {P} \biggr\} - {\partial \over \partial x_2} u_2 \biggl\{ {\partial \over \partial x_1} {P} \biggr \}+  {\partial \over \partial x_1} u_2  \biggl \{ {\partial \over \partial x_2} {P} \biggr\}  \\
& \qquad\qquad  -  {\partial \over \partial x_2} u_1 \biggl\{ {\partial \over \partial x_2} {P} \biggr\} - 2 \biggl( {\partial^2 \over \partial x_1^2} u_1 {P} +  {\partial^2 \over \partial x_1 \partial x_2} u_2 {P}  \biggr ) \\
& =    {\partial^2 \over \partial x_1^2}  ( u_1 {P} ) - {\partial \over \partial x_1} u_1 \biggl \{ {\partial \over \partial x_1} {P} \biggr \} \\
& \qquad + {\partial^2 \over \partial x_2^2}  ( u_1 {P} ) - {\partial \over \partial x_2} u_1 \biggl \{ {\partial \over \partial x_2} {P} \biggr \} \\
& \qquad \qquad +  {\partial \over \partial x_1} u_2  \biggl \{ {\partial \over \partial x_2} {P} \biggr \} - {\partial \over \partial x_2} u_2 \biggl \{ {\partial \over \partial x_1} {P} \biggr \}  \\
& =    {\partial \over \partial x_1} P {\partial u_1 \over \partial x_1} + {\partial \over \partial x_2} P {\partial u_1 \over \partial x_2} -  {\partial \over \partial x_1} P {\partial u_2 \over \partial x_2} +  {\partial \over \partial x_2} P {\partial u_2 \over \partial x_1} \\
& =   {\partial \over \partial x_1} P \biggl ( {\partial u_1 \over
\partial x_1} -  {\partial u_2 \over \partial x_2} \biggr ) + {\partial \over \partial x_2} P \biggl ( {\partial u_2 \over
\partial x_1} +  {\partial u_1 \over \partial x_2} \biggr).
\end{align*}

Similar calculations show that the momentum terms involving the
derivative of terms of the form $\rho u_i u_j$ all cancel. Thus we
have
\begin{multline*}
  {\partial \over \partial t} ( \rho u_1)  =  -\sum_{j=1}^2
{\partial
 \over \partial x_j} (\rho u_1 u_j)  -  {\partial P \over
\partial x_1}  \nonumber \\
 \qquad + {\tau \over 2} \biggl( {\partial \over \partial x_1} P
\biggl(  {\partial u_1 \over \partial x_1} -  {\partial u_2 \over
\partial x_2} \biggr) + {\partial \over \partial x_2} P \biggr(
{\partial u_2 \over \partial x_1} +  {\partial u_1 \over \partial
x_2} \biggr) \biggr),
\end{multline*}
which is the second of the Navier--Stokes equations~\eqref{nseq}
with $\mu=\tau/2$.

Thus we have demonstrated that, in performing an Ehrenfests' step
after free-flight we get, to the second-order, the Navier--Stokes
equations~\eqref{nseq} with coefficient of viscosity $\tau/2$. This
is remarkable, because it does not involve any particular form for
the collision integral in Boltzmann's equation~\eqref{be}, just
free-flight and equilibration.

\subsection{Decoupling time step and viscosity}

There is of course a difficulty in simulating a Navier--Stokes flow
where viscosity is given, with a numerical scheme in which the
viscosity is directly proportional to the time step. The free-flight
and equilibration scheme detailed above is such a scheme. We can
write the governing equation in the form
\begin{align*}
f_i(\xvec+\bvec_i \tau,t+\tau) & = f_i^*(\xvec,t) \\
& =  {1 \over 2} f_i(\xvec,t)+ {1 \over 2} f_i^{{\rm mir}}(\xvec,t),
\end{align*}
where $f_i^{{\rm mir}}(\xvec,t)=2f_i^*(\xvec,t)-f_i(\xvec,t)$. Thus,
after free-flight dynamics we move along a vector in the direction
of the \emph{mirror point}, $f^{{\rm mir}}$, which is the reflection
of $f$ in the quasiequilibrium manifold. With the BGK
collision~\eqref{bgkcoll} we move some part of the way along this
direction. This then suggests a more general numerical simulation
process
\begin{equation*}
f_i(\xvec+\bvec_i \tau,t+\tau) = (1-\beta) f_i(\xvec,t)+\beta
f_i^{{\rm mir}}(\xvec,t),\label{genproc}
\end{equation*}
where $\beta=\beta(\tau)$ may be chosen to satisfy a physically
relevant condition.

A choice of $\beta=1/2$ gives the Ehrenfests' step with viscosity
proportional to the time step $\Delta t=\tau$. For $\beta<1/2$ the
viscosity is even bigger. Hence, in these cases the time step gives
the lower boundary of viscosity we can realise. An important
development in LBM was the overrelaxation step, with
$\beta>1/2$~\cite{LBGK1,Higuera,LBGK2}. In this case the idea is
that the dynamics passes through the quasiequilibrium manifold so
that the next phase of free-flight would normally take us back
through the quasiequilibrium manifold. Now this method, commonly
called LBGK, is used for all $\beta$ from the stability interval
$\beta \in [0,1]$. For $\beta \to 1$ viscosity goes to zero. One
variant of LBGK is the so-called \emph{entropic} LBM (ELBM)
\cite{karlin06,karlin07,KGSBPRL} in which instead of a linear mirror
reflection $f \mapsto f^{{\rm mir}}$ an \emph{entropic involution}
$f \mapsto \tilde{f}$ is used, where $\tilde{f}= (1-\alpha) f+\alpha
f^*$. The number $\alpha=\alpha(f)$ is chosen so that the constant
entropy condition is satisfied: $S(f)=S(\tilde{f})$.

Both LBGK and ELBM decouple the viscosity parameter from the time
step. There are a number of other ways in which one can achieve the
same goal (see, e.g.,~\cite{BGJ,gorban06}). We do not concern
ourselves with this issue here but just remark that the essence is
to construct a numerical method from the dynamics
$\Theta_{-\tau/2}(f^*_M) \mapsto \Theta_{\tau/2}(f^*_M)$, then the
first-order term in $\tau$ is cancelled and one obtains an order
$\tau^2$ approximation to the Euler equations.

After discretization of velocity space, the additional space
discretization for LBM is not necessary in the following sense: the
restriction of the discrete-in-time and continuous-in-space LBM
chain of free-flights and collisions on a grid is exact, if this
grid is invariant with respect to parallel transitions on the
vectors $\bvec_i \tau$.

Unfortunately, as we will see in Sec.~\ref{numer} below, there are
instabilities in the simulation with LBGK and ELBM. This is because
the free-flight dynamics sometimes takes us too far (to be
understood in terms of entropy) from the quasiequilibrium manifold.
In this case we apply a single Ehrenfests' step and return to the
quasiequilibrium manifold. As you will see, this technique is
capable of stabilising the method. In order to retain an order
$\tau^2$ method (on average) we can only apply Ehrenfests' steps at
a bounded number of sites. Thus we fix a tolerance $\delta$ which
measures the entropy deviation $\Delta S=S(f^*)-S(f)$ from its
conditional maximum on the quasiequilibrium manifold, and then we
choose the $k$ (a fixed number) most distant points with $\Delta S >
\delta$ and return these to quasiequilibrium. If there is less than
$k$ such points, we choose to return all of them. We call $\Delta S$
\emph{nonequilibrium entropy}.

\subsection{Entropy control of non-entropic quasiequilibria}\label{entcontsec}

There are several ways to define the discrete quasiequilibria
$f^*_i$. One of them is by the postulating of moment conditions: the
moments $m(f^*)$ and their fluxes (moments of the next order,
usually) should coincide for the discrete quasiequilibrium and for
the corresponding continuous one (in this approach, ``continuous"
means ``genuine"). This is the approach used to derive the popular
polynomial quasiequilibria~\cite{succi01}. Another approach is based
on an entropy condition: the discrete system must have its own
thermodynamics and $H$-theorem, and the discrete quasiequilibrium
should be the conditional maximum of the discrete entropy.

We would like to apply Ehrenfests' stabilisation (as described at
the end of the previous section) for all sorts of quasiequilibria.
But this stabiliser requires the notion of entropy. In this section,
we demonstrate how to use this entropic stabiliser for non-entropic
quasiequilibria.

Let the discrete entropy have the standard form for ideal (perfect)
mixtures:
\begin{equation}\label{PerfEnt}
S(f)=-\sum_i f_i \ln\biggl(\frac{f_i}{W_i}\biggr).
\end{equation}
After the classical work of Zeldovich~\cite{Zeld}, this function is
recognized as a useful instrument for the analysis of kinetic
equations (especially in chemical kinetics~\cite{Kagan,YBGE}). For
applications in ELBM see~\cite{karlin99}.

If we define $f^*$ as the conditional entropy maximum~\eqref{argmax}
for given $M_j=\sum_k m_{jk} f_k$, then
\begin{equation*}
\ln f^*_k=\sum_j \mu_j m_{jk},
\end{equation*}
where $\mu_j(M)$ are the Lagrange multipliers (or ``potentials").
For this entropy and conditional equilibrium we find
\begin{equation}\label{DelS}
\Delta S= S(f^*)-S(f)=\sum_i f_i \ln\biggl(\frac{f_i}{f^*_i}\biggr)
\end{equation}
if $f$ and $f^*$ have the same moments, $m(f)=m(f^*)$.

The right hand side of~\eqref{DelS} is (minus) Kullback
entropy~\cite{Kull}. In thermodynamics, the Kullback entropy belongs
to the family of Massieu--Planck--Kramers functions (canonical or
grandcanonical potentials). The estimate of nonequilibrium entropy
$\Delta S$ can be performed for both entropic and non-entropic
quasiequilibria. Any quasiequilibrium (entropic or not) is the
conditional maximum of the Kullback entropy. The main difference
between the Kullback entropy~\eqref{DelS} of the form $-\sum_i f_i
\ln (f_i/f^*_{i})$ and the perfect entropy~\eqref{PerfEnt} is
dependence of the denominators $f^*_{i}$ on $M=m(f)$: $f^*=f^*_{M}$.
The perfect entropy~\eqref{PerfEnt} is a free-flight invariant, and
the Kullback entropy is not because of this dependence.

\section{Numerical Experiments} \label{numer}

To conclude this paper we report two numerical experiments conducted
to demonstrate the performance of the proposed Ehrenfests' step
stabilisation proposed in the previous section. The first test is a
$1$D shock tube and we are interested in testing the Ehrenfests'
regulariser on the LBGK and ELBM simulations for small (almost zero)
viscosity ($\nu \sim 10^{-9}$). We compare the LBGK simulation for
the popular polynomial quasiequilibria~\cite{succi01} and for
entropic quasiequilibria~\cite{karlin99}, as well as ELBM
simulation. In each case the scheme is supplemented by Ehrenfests'
steps in a small number $k$ sites with highest $\Delta S> \delta$.

The second test is the $2$D unsteady flow around a square-cylinder.
The unsteady flow around a square-cylinder has been widely
experimentally investigated in the literature (see, e.g.,
\cite{davis92,okajima82,vickery66}). We demonstrate that LBGK, with
the Ehrenfests' regularisation, is capable of quantitively capturing
the Strouhal--Reynolds relationship. The relationship is verified up
to $\mathrm{Re}=20000$ and compares well with Okajima's experimental
data~\cite{okajima82}.

\subsection{Shock tube}

The $1$D shock tube for a compressible isothermal fluid is a
standard benchmark test for hydrodynamic codes. We will fix the
kinematic viscosity of the fluid at $\nu=10^{-9}$ (essentially
zero). Our computational domain will be the interval $[0,1]$ and we
discretize this interval with $801$ uniformly spaced lattice sites.
We choose the initial density ratio as $1:2$ so that for $x\leq 400$
we set $\rho=1.0$, otherwise we set $\rho=0.5$.

In all of our simulations we use a lattice with spacing $h=1$, time
step $\tau=1$ and a discrete velocity set
$\{v_1,v_2,v_3\}:=\{0,-1,1\}$ so that the model consists of static,
left- and right-moving populations only. For a lattice site $x$, the
neighbouring lattice sites are $x+v_2$ and $x+v_3$. The governing
equations for LBGK are then
\begin{equation}\label{lbgkLB}
    f_i(x+v_i,t+1) =  f_i(x,t)+2 \beta(f_i^*(x,t)-f_i(x,t)),
\end{equation}
where the subscript $i$ denotes population (not lattice site number)
and $f_1$, $f_2$ and $f_3$ denote the static, left- and right-moving
populations, respectively.

The standard polynomial quasiequilibria~\cite{succi01} are
\begin{align*}
    f_1^* &= \frac{2\rho}{3}\biggl(1-\frac{3u^2}{2}\biggr),\\
    f_2^* &= \frac{\rho}{6}(1-3u+3u^2),\\
    f_3^* &= \frac{\rho}{6}(1+3u+3u^2),
\end{align*}
where we recall that
\begin{equation*}
  \rho := \sum_i f_i,\qquad \rho u:=\sum_i v_i f_i.
\end{equation*}

For entropic quasiequilibria the entropy is $S=-H$, with
\begin{equation*}
 H = f_1 \log(f_1/4)+f_2\log(f_2)+f_3 \log(f_3),
\end{equation*}
(see, e.g.,~\cite{karlin99}). For this entropy the quasiequilibrium
is available explicitly:
\begin{align*}
  f_1^{*} &= \frac{2 \rho}{3} \bigl(2-\sqrt{1+3 u^2}\bigr),\\
  f_2^{*} &= \frac{\rho}{6} \bigl((3u-1)+2\sqrt{1+3 u^2} \bigr),\\
  f_3^{*} &= -\frac{\rho}{6} \bigl((3u+1)-2\sqrt{1+3 u^2} \bigr).
\end{align*}

For our realisation of the Ehrenfests' regularisation, which is
intended to keep states uniformly close to the quasiequilibrium
manifold, we monitor nonequilibrium entropy $\Delta S$ at every
lattice site throughout the simulation. If a pre-specified threshold
value $\delta$ is exceeded, then an Ehrenfests' step is taken at the
corresponding site. Now, the governing LBGK equations become:
\begin{equation}\label{ESLB}
    f_i(x+v_i ,t+1) = \left\{
    \begin{aligned}
        &f_i(x,t)+2\beta(f_i^*(x,t)-f_i(x,t)),&& \text{$\Delta S \leq \delta$,} \\
        &f_i(x,t),&\quad& \text{otherwise.}
    \end{aligned}\right.
\end{equation}
We select the $k$ sites with highest $\Delta S>\delta$ so that the
Ehrenfests' steps are not allowed to degrade the accuracy of LBGK.

For ELBM, entropic quasiequilibria are always employed and the
governing equation is
\begin{equation}\label{elbmLB}
    f_i(x+v_i,t+1) =  f_i(x,t)+\alpha\beta(f_i^*(x,t)-f_i(x,t)).
\end{equation}
This equation differs from LBGK by the introduction of a parameter
$\alpha$ which is selected to satisfy the constant entropy
condition:
\begin{equation*}
 S(f+\alpha(f^*-f))=S(f).
\end{equation*}
This is a nonlinear equation for $\alpha$ which we solve, using the
bisection method, to an accuracy of $10^{-15}$ (see~\cite{Rob2} for
further details of the implementation). Supplementing ELBM with
Ehrenfests' steps is the same as for LBGK.

\begin{figure}
\begin{centering}
\includegraphics[width=11cm]{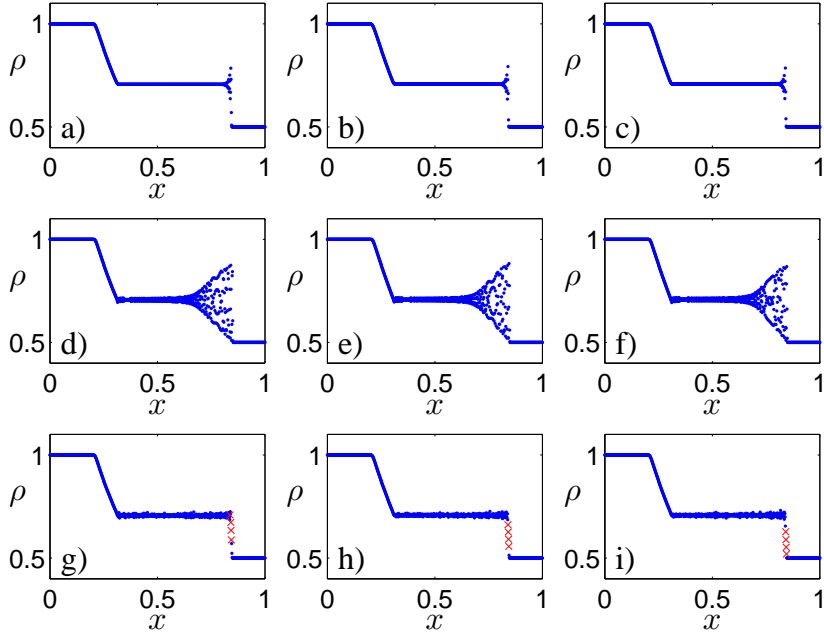}\\
\caption{Density and velocity profile of the 1:2 isothermal shock
tube simulation after $400$ time steps using (a) LBGK with
polynomial quasiequilibria~\eqref{lbgkLB} [$\nu=3.3333\times
10^{-2}$]; (b) LBGK with entropic quasiequilibria~\eqref{lbgkLB}
[$\nu=3.3333\times 10^{-2}$]; (c) ELBM~\eqref{elbmLB}
[$\nu=3.3333\times 10^{-2}$]; (d) LBGK with polynomial
quasiequilibria~\eqref{lbgkLB} [$\nu=10^{-9}$]; (e) LBGK with
entropic quasiequilibria~\eqref{lbgkLB} [$\nu=10^{-9}$]; (f)
ELBM~\eqref{elbmLB} [$\nu=10^{-9}$]; (g) LBGK with polynomial
quasiequilibria and Ehrenfests' steps~\eqref{ESLB} [$\nu=10^{-9}$,
$(k,\delta)=(4,10^{-4})$]; (h) LBGK with entropic quasiequilibria
and Ehrenfests' steps~\eqref{ESLB} [$\nu=10^{-9}$,
$(k,\delta)=(4,10^{-4})$]; (i) ELBM with Ehrenfests' steps
[$\nu=10^{-9}$, $(k,\delta)=(4,10^{-4})$]. Sites where Ehrenfests'
steps are employed are indicated by crosses.\label{shocktube}}
\end{centering}
\end{figure}

We observe that the Ehrenfests' stabilisation recipe is capable of
subduing spurious post-shock oscillations whereas LBGK fails in this
respect (Fig.~\ref{shocktube}). In the example we have considered a
fixed tolerance of $(k,\delta)=(4,10^{-4})$. Of course, we note also
that the smaller the tolerance $\delta$ the more smoothing we have
of the shock. Therefore we reiterate that it is important for
Ehrenfests' steps to be employed at only a small proportion of the
sites.

We do not detect any advantage of using ELBM over LBGK with entropic
quasiequilibria for this example. However, there appears to be some
gain in employing entropic rather than polynomial quasiequilibria.
We observe that the post-shock region for the unregularised LBGK
simulations is more oscillatory when polynomial quasiequilibria are
used. In Fig.~\ref{shocktube} we have also included a panel with the
simulation resulting from a much higher viscosity ($\nu=3.3333\times
10^{-2}$). Here, we observe no appreciable differences in the
results of LBGK and ELBM.

\subsection{Flow around a square-cylinder}

Our second test is the $2$D unsteady flow around a square-cylinder.
The realisation of LBGK that we use will employ a uniform $9$-speed
square lattice with discrete velocities
\begin{equation*}
    \bvec_i = \left\{
    \begin{aligned}
        &0,&\quad& \text{$i=0$,} \\
        & \biggl(\cos\Bigl((i-1)\frac{\pi}{2}\Bigr) ,\sin\Bigl((i-1)\frac{\pi}{2}\Bigr)\biggr) ,&& \text{$i=1,2,3,4$,}\\
        & \sqrt{2} \biggl(\cos\Bigl((i-5)\frac{\pi}{2}+\frac{\pi}{4}\Bigr),\sin\Bigl((i-5)\frac{\pi}{2}+\frac{\pi}{4}\Bigr)\biggr) ,&& \text{$i=5,6,7,8$.}\\
    \end{aligned}\right.
\end{equation*}
The numbering $f_0$, $f_1, \dotsc, f_8$  are for the static, east-,
north-, west-, south-, northeast-, northwest-, southwest- and
southeast-moving populations, respectively. Here, we select entropic
quasiequilibria by maximising the entropy functional
\begin{equation*}
S(f) = -\sum_i f_i \log{\Bigl(\frac{f_i}{W_i}\Bigr)},
\end{equation*}
subject to the constraints of conservation of mass and
momentum~\cite{ansumali03}:
\begin{equation*}
  f_i^* = \rho W_i \prod_{j=1}^2
  \Bigl(2-\sqrt{1+3u_j^2}\Bigr)\Biggl( \frac{2u_j+\sqrt{1+3u_j^2}}{1-u_j} \Biggr)^{v_{i,j}}
\end{equation*}
Here, the \emph{lattice weights}, $W_i$, are given lattice-specific
constants: $W_0=4/9$, $W_{1,2,3,4}=1/9$ and $W_{5,6,7,8}=1/36$. As
is usual, the macroscopic variables are given by the expressions
\begin{equation*}
  \rho := \sum_i f_i,\qquad \rho\uvec  := \sum_i \bvec_i f_i.
\end{equation*}

The computational set up for the flow is as follows. A
square-cylinder of side length $L$, initially at rest, is emersed in
a constant flow in a rectangular channel of length $30L$ and height
$25L$. The cylinder is place on the centre line in the $y$-direction
resulting in a blockage ratio of $4$\%. The centre of the cylinder
is placed at a distance $10.5L$ from the inlet. The free-stream
fluid velocity is fixed at $(u_\infty,v_\infty)=(0.05,0)$ (in
lattice units) for all simulations.

On the north and south channel walls a free-slip boundary condition
is imposed (see, e.g.,~\cite{succi01}). At the inlet, the inward
pointing velocities are replaced with their quasiequilibrium values
corresponding to the free-stream fluid velocity. At the outlet, the
inward pointing velocities are replaced with their associated
quasiequilibrium values corresponding to the velocity and density of
the penultimate row of the lattice. Some care should to be taken
with the boundary conditions on the cylinder, but for more
information on these the reader may consult,
e.g.,~\cite{ansumali02,ansumali04}.

\subsubsection{Strouhal--Reynolds relationship}

As a test of the Ehrenfests' regularisation, a series
of simulations, all with characteristic length fixed at $L=20$, were
conducted over a range of Reynolds numbers
The parameter pair $(k,\delta)$, which control the Ehrenfests' steps
tolerances, are fixed at $(L/2,10^{-3})$.

We are interested in computing the Strouhal--Reynolds relationship.
The Strouhal number $\mathrm{St}$ is a dimensionless measure of the
vortex shedding frequency in the wake of one side of the cylinder:
\begin{equation*}
  \mathrm{St} = \frac{L f_\omega}{u_\infty},
\end{equation*}
where $f_\omega$ is the shedding frequency.

For our computational set up, the vortex shedding frequency is
computed using the following algorithmic technique. Firstly, the
$x$-component of velocity is recorded during the simulation over
$t_\mathrm{max} = 1250L/u_\infty$ time steps. The monitoring points
is positioned at coordinates $(4L,-2L)$ (assuming the origin is at
the centre of the cylinder). Next, the dominant frequency is
extracted from the final $25$\% of the signal using the discrete
Fourier transform. The monitoring point is purposefully placed
sufficiently downstream and away from the centre line so that only
the influence of one side of the cylinder is recorded.

\begin{figure}[t]
\begin{centering}
    \includegraphics[width=9.0cm]{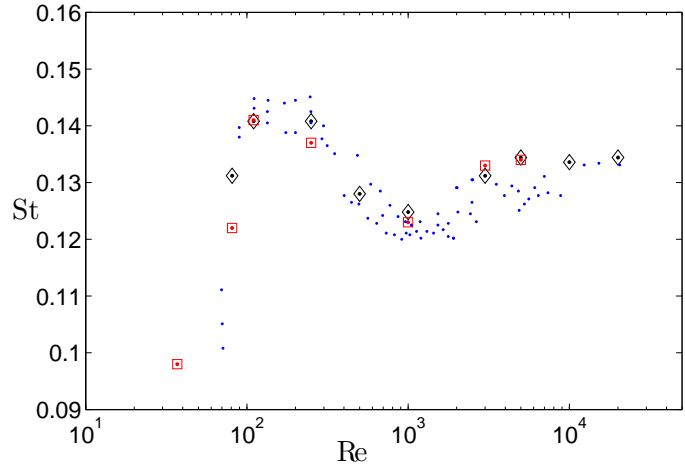}
    \caption{Variation of Strouhal number as a function of Reynolds. Dots are Okajima's
    experimental data~\cite{okajima82} (the data has been digitally
extracted from the original paper). Diamonds are the Ehrenfests'
regularisation of LBGK and the squares are the ELBM simulation
from~\cite{ansumali04}.}\label{sqcy}
\end{centering}
\end{figure}

The computed Strouhal--Reynolds relationship using the Ehrenfests'
regularisation of LBGK is shown in Fig.~\ref{sqcy}. The simulation
compares well with Okajima's data from wind tunnel and water tank
experiment~\cite{okajima82}. The present simulation extends previous
LBM studies of this problem~\cite{ansumali04,baskar04} which have
been able to quantitively captured the relationship up to
$\mathrm{Re}\sim 1000$. Fig.~\ref{sqcy} also shows the ELBM
simulation results from~\cite{ansumali04}. Furthermore, the
computational domain was fixed for all the present computations,
with the smallest value of the kinematic viscosity attained being
$\nu = 5\times 10^{-5}$ at $\mathrm{Re}=20000$. It is worth
mentioning that, for this characteristic length, LBGK exhibits
numerical divergence at around $\mathrm{Re}=1000$. We estimate that,
for the present set up, the computational domain would require at
least $\sim 10^7$ lattice sites for the kinematic viscosity to be
large enough for LBGK to converge at $\mathrm{Re}=20000$. This is
compared with $\sim 10^5$ sites for the present simulation.

\begin{acknowledgements}
This work is supported by Engineering and Physical Sciences Research
Council (EPSRC) grant number GR/S95572/01. The authors acknowledge
I.~V.~Karlin for important discussions, S.~Chikatamarla for kindly
providing the digitally extracted data used in Fig.~\ref{sqcy} and
anonymous referee for valuable comments.
\end{acknowledgements}

\appendix
\section{Moments of the quasiequilibrium distribution}

In this appendix we calculate the moments of the distribution $f^*$
so as to keep the presentation in the main body of the paper clean.
All integrals are over the whole velocity space. Firstly, we have by
definition
\begin{equation*}
 \int f^* \,\mathrm{d}\bvec  =  \rho, \qquad \int v_i f^*
\,\mathrm{d}\bvec  =  \rho u_i, \qquad \int \bvec^2 f^*
\,\mathrm{d}\bvec =  2 E = 2 P + \rho \uvec^2.
\end{equation*}

Next, we have
\begin{equation}\label{vivj}
\int v_i v_j f^* \,\mathrm{d}\bvec =  \int (v_i-u_i)(v_j-u_j) f^*
\,\mathrm{d}\bvec + \rho u_i u_j.
\end{equation}
Now, using the identity
\begin{equation*}
 \int \alpha \beta \mathrm{e}^{-(\alpha^2+\beta^2)}\,\mathrm{d}\alpha \mathrm{d}\beta  =
  0,
\end{equation*}
it follows by a change of variables that, for $i\neq j$,
\begin{equation*}
 \int (v_j-u_j)(v_i-u_i) f^* \,\mathrm{d}\bvec =  0.
\end{equation*}
Further, it follows from the identity
\begin{equation*}
  \int \alpha^2 \mathrm{e}^{-(\alpha^2+\beta^2)} \,\mathrm{d}\alpha \mathrm{d}\beta =
  \int \beta^2 \mathrm{e}^{-(\alpha^2+\beta^2)} \,\mathrm{d}\alpha \mathrm{d}\beta
\end{equation*}
that
\begin{equation*}
\begin{split}
 \int (v_j-u_j)^2 f^* \,\mathrm{d}\bvec &= \frac{1}{2}  \int (\bvec-\uvec)^2 f^* \,\mathrm{d}\bvec  \\
&= \frac{1}{2}  \int (\bvec^2+\uvec^2 - 2v_1 u_1 -2v_2 u_2 ) f^* \,\mathrm{d}\bvec  \\
&= \frac{1}{2} ( {2 P} + \rho \uvec^2 + \rho \uvec^2 - 2 \rho u_1^2 - 2\rho u_2^2 )  \\
&=  P.
\end{split}
\end{equation*}
Hence, from~\eqref{vivj}, we have
\begin{equation}\label{ivivj}
\int v_i v_j f^* \,\mathrm{d}\bvec =  \delta_{i,j} {P}+\rho u_i u_j.
\end{equation}

Finally, similar calculations provide us with
\begin{equation}\label{ivlvivj}
\int v_\ell v_i v_j f^* \,\mathrm{d}\bvec  = (\delta_{i,j} u_\ell +
\delta_{\ell,j} u_i + \delta_{\ell,i} u_j){P} + \rho u_\ell u_i u_j.
\end{equation}

\bibliographystyle{plain}

\begin{thebibliography}{10}

\bibitem{ansumali02}
S.~Ansumali and I.~V.~Karlin.
\newblock Kinetic boundary conditions in the lattice {B}oltzmann method.
\newblock {\em Phys. Rev. E}, 66(2):026311, 2002.

\bibitem{ansumali03}
S.~Ansumali S, I.~V.~Karlin, H.C. Ottinger,
\newblock Minimal entropic kinetic models for hydrodynamics
\newblock {\em Europhys. Let.} 63 (6): 798-804. 2003.

\bibitem{ansumali04}
S.~Ansumali, S.~S.~Chikatamarla, C.~E.~Frouzakis, and K.~Boulouchos.
\newblock Entropic lattice {B}oltzmann simulation of the flow past
  square-cylinder.
\newblock {\em Int. J. Mod. Phys. C}, 15:435--445, 2004.

\bibitem{baskar04}
G.~Baskar and V.~Babu.
\newblock Simulation of the unsteady flow around rectangular cylinders using
  the {ISLB} method.
\newblock In {\em 34th AIAA Fluid Dynamics Conference and Exhibit}, pages
  AIAA--2004--2651, 2004.

\bibitem{bgk} P.~L.~Bhatnagar, E.~P.~Gross, and M.~Krook.
\newblock A model for collision processes in gases I. Small amplitude processes in charged and neutral one-component systems.
\newblock {\em Phys. Rev.} 94(3):511--525, 1954.

\bibitem{Rob2}
R.~A. Brownlee, A.~N. Gorban, and J.~Levesley.
\newblock Stabilisation of the lattice-{B}oltzmann method using the
  {E}hrenfests' coarse-graining.
\newblock {\em cond-mat/0605359}, 2006.

\bibitem{Rob}
R.~A. Brownlee, A.~N. Gorban, and J.~Levesley.
\newblock Stabilisation of the lattice-{B}oltzmann method using the
  {E}hrenfests' coarse-graining.
\newblock {\em Phys. Rev. E}, 74:037703, 2006.

\bibitem{BGJPreprint} R.~A.~Brownlee, A.~N.~Gorban, and J.~Levesley, Stability and
stabilization of the lattice Boltzmann method: Magic steps and
salvation operations. {\em cond-mat/0611444}, 2006.

\bibitem{BGJ} R.~A.~Brownlee, A.~N.~Gorban, and J.~Levesley, Stability and
stabilization of the lattice Boltzmann method.
\newblock {\em Phys. Rev. E}, to appear.

\bibitem{LBGK1} H.~Chen, S.~Chen, W.~Matthaeus, Recovery of the Navier--Stokes
equation using a lattice--gas Boltzmann Method {\it Phys. Rev. A} 45
(1992), R5339--R5342.

\bibitem{davis92}
R.~W. Davis and E.~F. Moore.
\newblock A numerical study of vortex shedding from rectangles.
\newblock {\em J. Fluid Mech.}, 116:475--506, 1982.

\bibitem{ehrenfest11}
P.~Ehrenfest and T.~Ehrenfest.
\newblock {\em The conceptual foundations of the statistical approach in
  mechanics}.
\newblock Dover Publications Inc., New York, 1990.

\bibitem{gorban06}
A.~N. Gorban.
\newblock Basic types of coarse-graining.
\newblock In: A.~N. Gorban, N.~Kazantzis, I.~G. Kevrekidis, H.-C. \"{O}ttinger,
  and C.~Theodoropoulos, editors, {\em Model Reduction and Coarse-Graining
  Approaches for Multiscale Phenomena}, pages 117--176. Springer,
  Berlin-Heidelberg-New York, 2006.
\newblock cond-mat/0602024.

\bibitem{Kagan} A.~Gorban, B.~Kaganovich, S.~Filippov, A.~Keiko, V.~Shamansky,
I.~Shirkalin, {\em Thermodynamic Equilibria and Extrema: Analysis of
Attainability Regions and Partial Equilibrium}, Springer, Berlin,
Heidelberg, New York,   2006 (in press).

\bibitem{GorKar}
A.~N.~Gorban and I.~V.~Karlin.
\newblock {\em Invariant manifolds for physical and chemical kinetics}, volume
  660 of Lect. Notes Phys.
\newblock Springer, Berlin-Heidelberg-New York, 2005.

\bibitem{GKOeTPRE2001}
A.~N. Gorban, I.~V. Karlin, H.~C. \"{O}ttinger, and L.~L.
Tatarinova.
\newblock Ehrenfest's argument extended to a formalism of nonequilibrium
  thermodynamics.
\newblock {\em Phys. Rev. E}, 62:066124, 2001.

\bibitem{Higuera}
F.~Higuera, S.~Succi, and R.~Benzi.
\newblock Lattice gas -- dynamics with enhanced collisions.
\newblock {\em Europhys. Lett.}, 9:345--349, 1989.

\bibitem{karlin06}
I.~V.~Karlin, S.~Ansumali, C.~E.~Frouzakis, and S.~S.~Chikatamarla,
\newblock Elements of the lattice Boltzmann method I: Linear advection
equation.
\newblock {\em Commun. Comput. Phys.}, 1 (2006), 616--655.

\bibitem{karlin07}I.~V.~Karlin, S.~S.~Chikatamarla and S.~Ansumali.
\newblock Elements of the lattice Boltzmann method II: Kinetics and hydrodynamics in one
dimension.
\newblock {\em Commun. Comput. Phys.}, 2 (2007), 196--238.

\bibitem{karlin99}
I.~V. Karlin, A.~Ferrante, and H.~C. \"{O}ttinger.
\newblock Perfect entropy functions of the lattice {B}oltzmann method.
\newblock {\em Europhys. Lett.}, 47:182--188, 1999.

\bibitem{KGSBPRL}
I.~V. Karlin, A.~N. Gorban, S.~Succi, and V.~Boffi.
\newblock Maximum entropy principle for lattice kinetic equations.
\newblock {\em Phys. Rev. Lett.}, 81:6--9, 1998.

\bibitem{Kull}S.~Kullback, {\em Information theory and statistics}, Wiley, New York, 1959.

\bibitem{okajima82}
A.~Okajima.
\newblock Strouhal numbers of rectangular cylinders.
\newblock {\em J. Fluid Mech.}, 123:379--398, 1982.

\bibitem{shanhe}
X. Shan and X. He
\newblock Discretization of the velocity space in the solution of the Boltzmann equation
\newblock {\em Phys. Rev. Lett.}, 80:65--68, 1998.

\bibitem{succi01}
S.~Succi.
\newblock {\em The lattice {B}oltzmann equation for fluid dynamics and beyond}.
\newblock OUP, New York, 2001.

\bibitem{LBGK2}Y.~H.~Qian, D.~d'Humieres, P.~Lallemand,
Lattice BGK models for Navier--Stokes equation, {\it Europhys.
Lett.} 17, (1992), 479--484.

\bibitem{vickery66}
B.~J. Vickery,
\newblock Fluctuating lift and drag on a long cylinder of square cross-section
  in a smooth and in a turbulent stream.
\newblock {\em J. Fluid Mech.}, 25:481--494, 1966.

\bibitem{YBGE}G.~S.~Yablonskii, V.~I.~Bykov, A.~N.~Gorban, and V.~I.~Elokhin,  Kinetic
Models of Catalytic Reactions (Series ``Comprehensive Chemical
Kinetics," V.32, ed. by R.~G.~Compton), Elsevier, Amsterdam, 1991.

\bibitem{Zeld}Y.~B.~Zeldovich, Proof of the Uniqueness of the Solution of the
Equations of the Law of Mass Action, In: {\it Selected Works of
Yakov Borisovich Zeldovich,} Vol. 1, J.~P.~Ostriker (Ed.), Princeton
University Press, Princeton, USA, 1996, 144--148.

\end{thebibliography}

\end{document}